\documentclass[11pt,twoside]{article}

\usepackage{asp2006}
\usepackage{epsf}
\usepackage{psfig}
\usepackage{lscape}

\begin{document}
\title{Near-Infrared spectral diagnostics for unresolved stellar 
population galaxies}   
\author{M. Cesetti$^{1,2}$, V. D. Ivanov$^{1}$, L. Morelli$^{2}$}   
\affil{$^{1}$European Southern Observatory, Av. Alonso de C\'ordoba 
3107, Casilla 19, Santiago 19001, 
Chile, \\
$^{2}$Dipartimento di Astronomia, Universit\`a di Padova, vicolo 
dell'Osservatorio 2, I-35122 
Padova, Italy }    

\begin{abstract} 
We want to develop spectral diagnostics of stellar 
populations in the near-infrared (NIR), for unresolved stellar 
populations. We created a semi-empirical population model 
and we compare the model output with the observed spectra 
of a sample of elliptical and bulge-dominated galaxies that have reliable 
Lick-indices from literature to test if the correlation between Mg2 and 
CO 1.62 $\mu$m remains valid in galaxies and to calibrate it as an 
abundance indicator. We find that (i) there are no significant correlations 
between any NIR feature and 
the optical Mg2; (ii) the CaI, NaI and CO trace the $\alpha$-enhancement; 
and (iii) the NIR absorption features are not influenced by the galaxy’s age.
\end{abstract}


\section*{Introduction}   

Accurate abundance determinations through optical spectroscopy are 
impossible for heavily obscured evolved stellar populations such 
as the dusty spheroids or bulge-embedded globular clusters 
\citep[cf.][]{Ste04}. Furthermore, the visible may not be representative 
of the older stellar populations in the galaxies. There are 
promising NIR counterparts of some important visible diagnostics: 
i.e. in stars, the optical age indicator H$\beta$ correlates well 
with Br$\gamma$ while the metallicity indicators Mgb and Mg2 
correlate with MgI\,1.50\,$\mu$m and CO\,1.62\,$\mu$m, respectively. 
We use a sample of elliptical and bulge-dominate galaxies that have 
reliable Lick-indices from \citet{Ben93} to test if these correlations 
remains valid in galaxies and to calibrate the IR Mg features as 
abundance indicators.

\section*{The Model}   

\begin{figure}
\plottwo{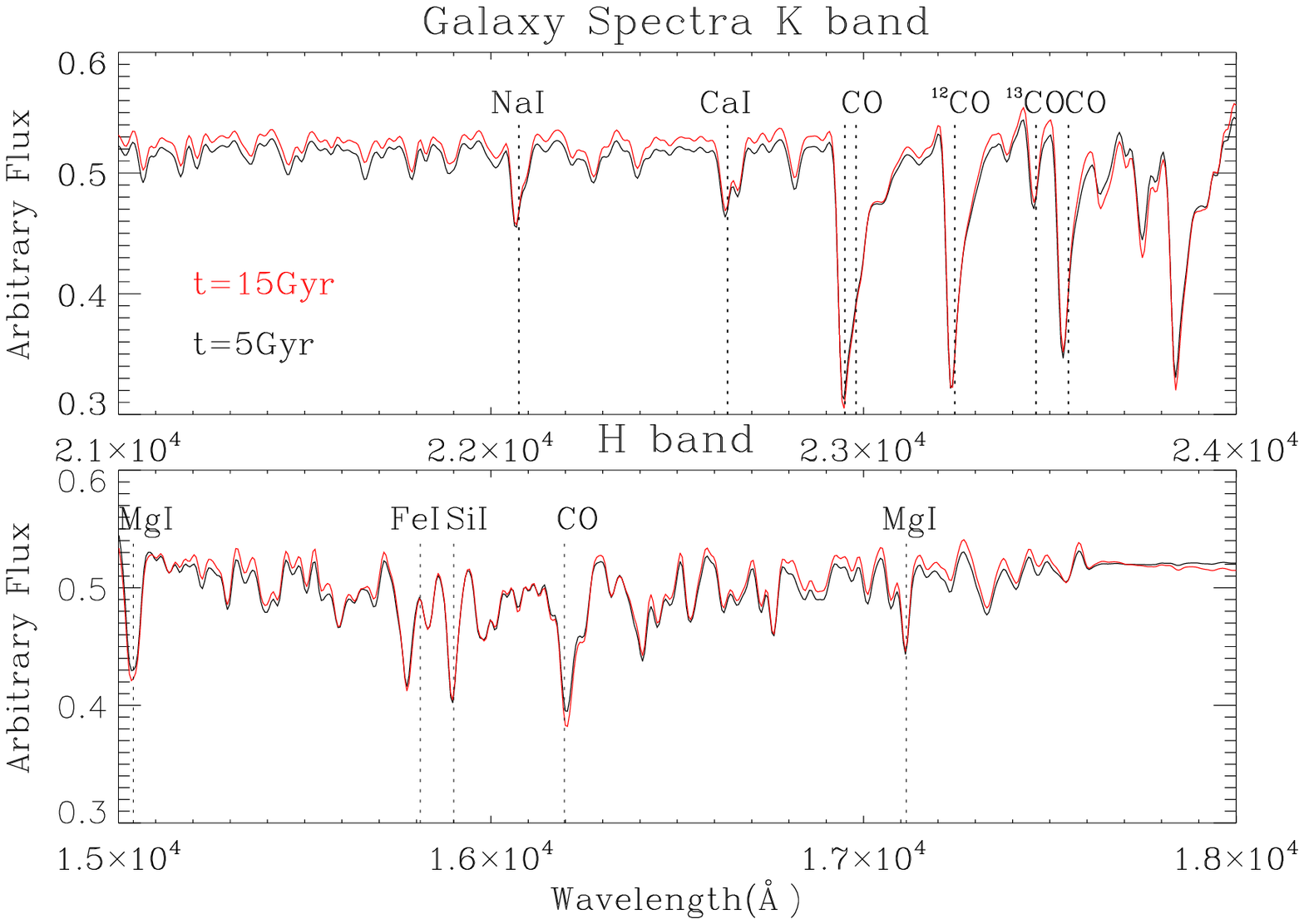}{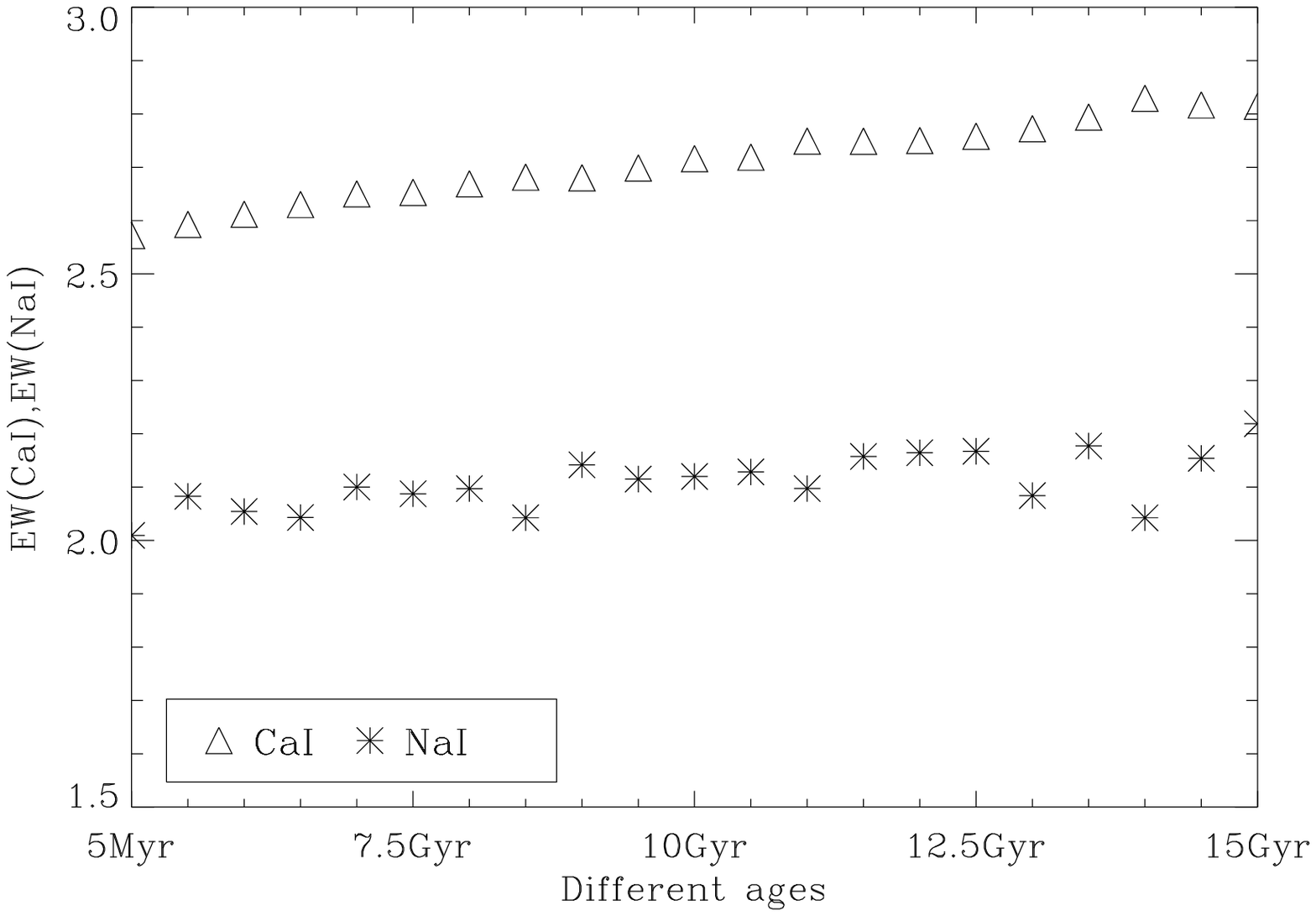}
\caption{Age variation. \textit{Left panel}: Synthetic spectra for
5\,Gyr (black or solid line) and 15\,Gyr (red or dashed line) old galaxies. 
\textit{Right panel}: Behavior of CaI and NaI EW in our synthetic 
spectra for a range of ages.}
\label{AgeVar}
\end{figure}

We created a NIR population synthesis model based on single age, 
single metallicity stellar populations (SSPs) that describes the 
parameters of the entire population with a single isochrone. The 
stellar evolution models are from \citet{Pie04} and the IMF is from 
\citet{Kro93}. The HK IR stellar library of \citet{Iva04} was used. 
In general, our model predicts weaker NIR features than the 
observed ones in elliptical galaxies, probably due to the higher 
$\alpha$-to-Iron ratio in the ellipticals, with respect to the Solar 
neighborhood from where most of our library stars come from. We 
presented here our NIR (1.5-2.5\,$\mu$m) synthetic spectra, and 
we examine how they are affected by changes in age and IMF
(Fig.\,\ref{AgeVar}). The model predicts that the IR features are 
only weakly IMF and age dependent. 

\section*{Correlations between the Optical and the NIR features}

\begin{figure}
\plottwo{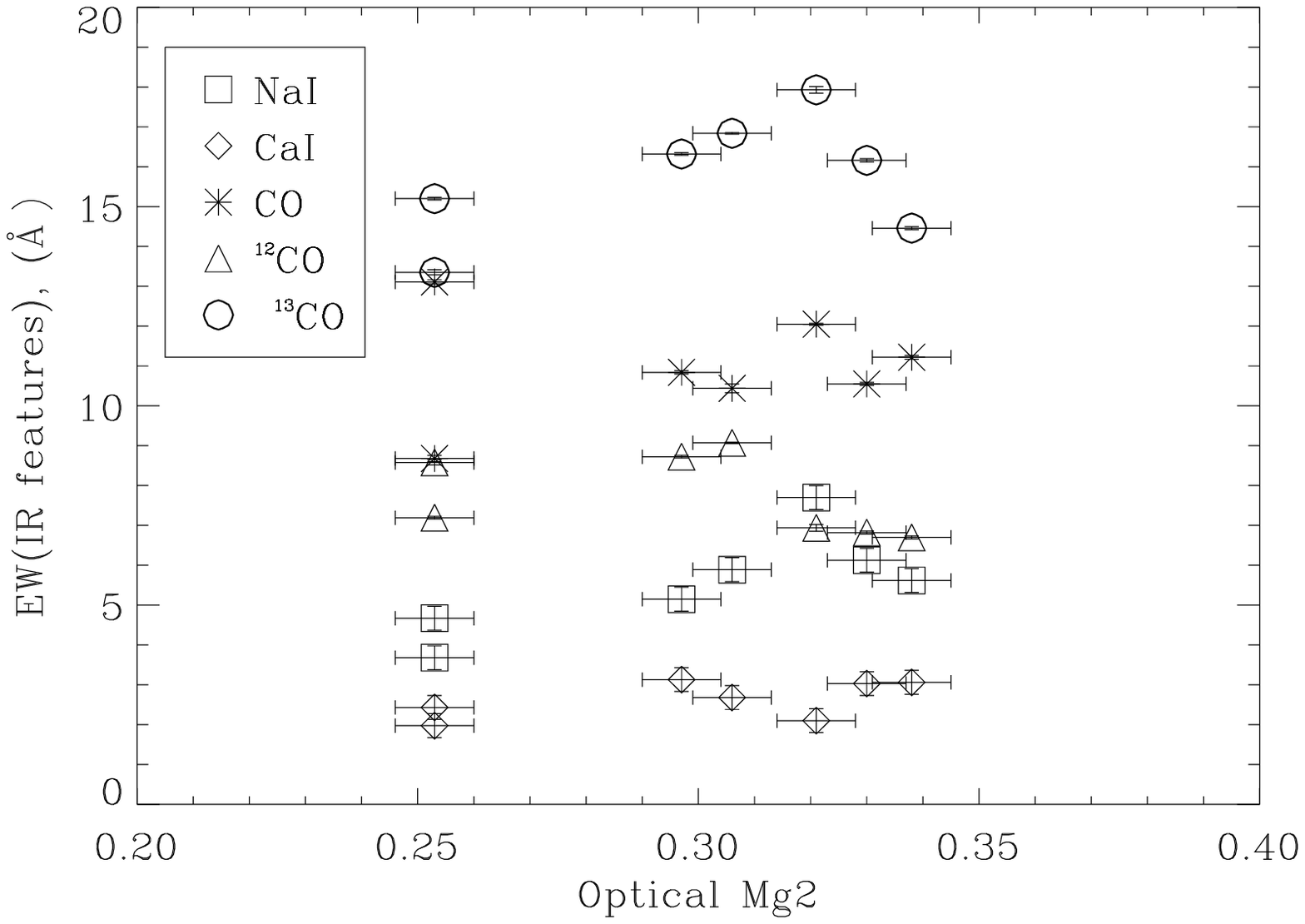}{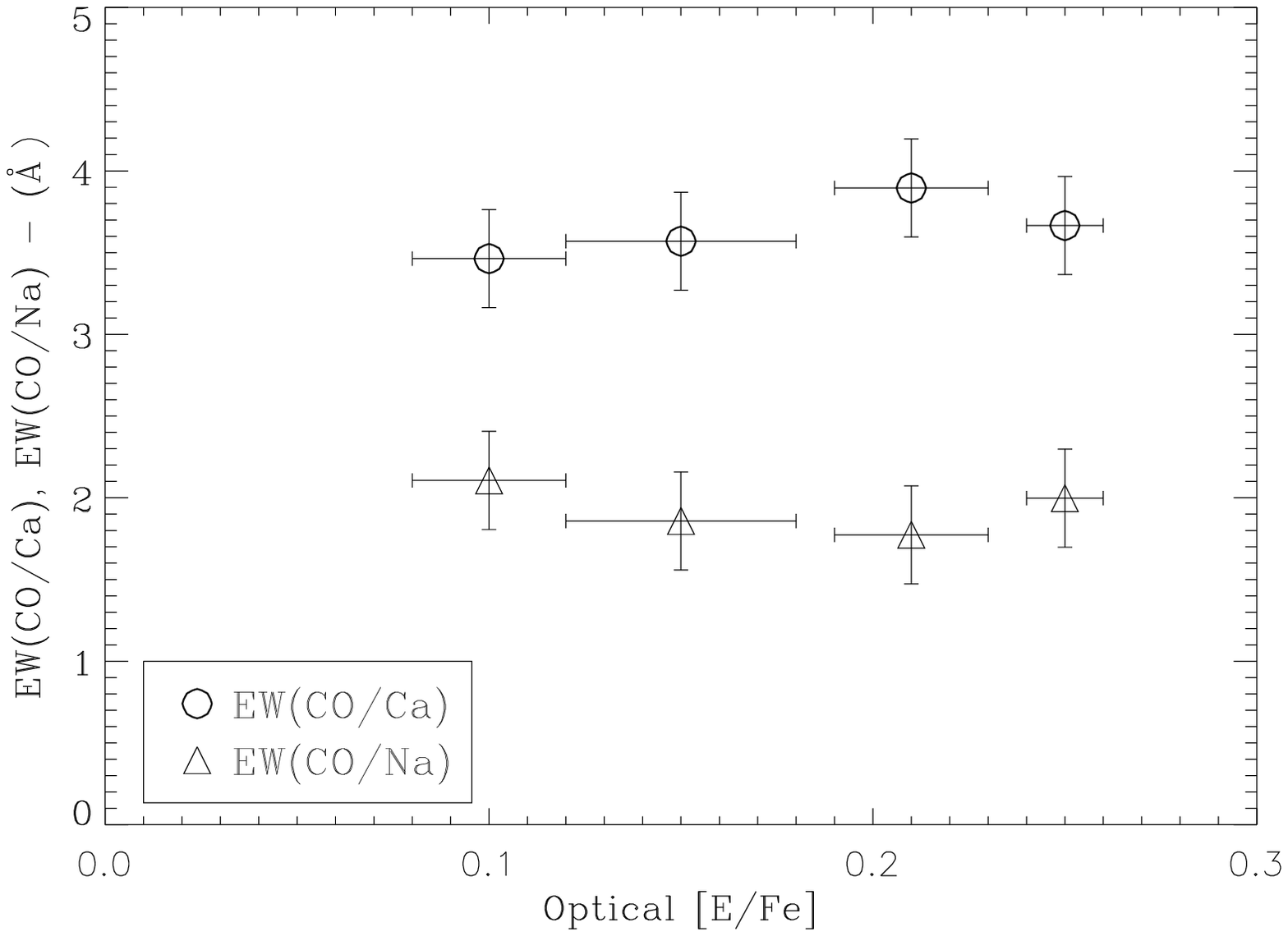}
\caption{Optical-to-NIR index correlations. \textit{Left panel}: 
Mg2 vs. some NIR features for a sample of elliptical and S0 galaxies. 
\textit{Right panel}: Optical $\alpha$-enhancement versus IR CO-to-CaI 
and CO-to-NaI ratio.}
\label{IndexCorr}
\end{figure}

We found no significant trend between the optical Mg2 and the NIR 
features in galaxies (Fig.\,\ref{IndexCorr}). The IR CO-to-atomic 
features ratio vs. the $\alpha$-elements enrichment in optical for 
4 observed galaxies is consistent with a constant. Provisionally, 
we can conclude that the CaI and NaI features trace the $\alpha$ 
enhancement, same as the CO.




\end{document}